\documentclass[11pt]{article}
\usepackage{latexsym}
\usepackage{graphicx}
\usepackage{amsfonts,amssymb}

\begin{document}

\title{Effects of Weight on Structure and Dynamics in Complex
Networks}% Force line breaks with \\

\author{Menghui Li$^1$, Ying Fan$^{1}$\footnote{Author for correspondence:
yfan@bnu.edu.cn}, Dahui Wang$^1$, Na Liu$^1$, Daqing Li$^1$,\\ Jinshan Wu$^{2}$, Zengru Di$^1$\\
\\ 1. Department of Systems Science, School of Management,\\
Beijing Normal University, Beijing 100875, P.R.China
\\2. Department of Physics \& Astronomy, University of British
Columbia,\\ Vancouver, B.C. Canada, V6T 1Z1.\\
}

\maketitle
%\affiliation{}%

\begin{abstract}Link weight is crucial in weighted complex networks. It provides additional
dimension for describing and adjusting the properties of networks.
The topological role of weight is studied by the effects of random
redistribution of link weights based on regular network with
initial homogeneous weight. The small world effect emerges due to
the weight randomization. Its effects on the dynamical systems
coupled by weighted networks are also investigated. Randomization
of weight can increase the transition temperature in Ising model
and enhance the ability of synchronization of chaotic systems
dramatically.

\end{abstract}


%\section{Introduction}
Network analysis is now widely used to describe the relationship
and collective behavior in many fields\cite{Review1, Review2}. A
binary network has a set of vertices and a set of edges which
represent the relationships between any two vertices. Obviously,
the edge only represents the presence or absence of interaction.
It will be a strong limitation when such an approach is used to
describe relations with different strength or having more levels.
In fact, the interaction strength usually plays an important role
in many real networks. For example, the number of passengers or
flights between any two airports in airport
networks\cite{W.Li,Barrat1,Bagler}, the closeness of any two
scientists in scientific collaboration
networks\cite{Newman1,Newman2,Fan,Li}, the reaction rates in
metabolic networks\cite{Nature} are all crucial in the
characterization of the corresponding networks. So it is necessary
to take the link weight into account and to study weighted
networks. Recently more and more studies in complex networks focus
on the weighted networks. The problems involve the definition of
weight and other quantities which characterize the weighted
networks\cite{Analysis,BBPV,Li}, the empirical studies about its
statistical properties\cite{Almaas,Bianconi,Jezewski,Fan},
evolving models\cite{Yook,BBV,Antal,Dorogovtsev,WangW}, and
transportation or other dynamics on weighted
networks\cite{Goh,KIGoh,ZhouT}.

Weight, a quantity reflects the strength of interaction, gives
more information about networks besides its topology dominated by
links. Weight provides additional dimension to describe and to
adjust network properties. However, how important is the weight,
or what significant changes on network structures are induced when
weight is changed? This question is related to the role of weight.
It should be a fundamental problem in the study of weighted
networks. But it has not been investigated deeply in previous
studies.

For weighted networks, the redistribution of weight on links
provide another way to adjust network structure besides the change
of links between vertices. The role of weight could be studied
both by its effects on network structures and its effects on
dynamical processes taking place on networks. The effects of
weight on network structures can be investigated on single vertex
statistics, such as degree and clustering coefficient, and global
properties such as distance, betweenness and especially community
structure. The effects of weight on dynamics can be examined by
the collective behaviors of dynamical systems, such as the
synchronization of oscillators or chaos, the phase transition of
spin system, the coherent oscillations of excitable systems, the
spread of an infectious disease, the propagation of information,
and so on. Except we consider the difference of above properties
between unweighted and weighted networks, an efficient way to
study the effects of weight is to consider the difference after
disturbing the weight distribution. We have introduced the way to
disturb the weight distribution and investigated its effects on
network structures including single vertex statistics,
distance\cite{Li}, and community structure\cite{Fan2} in several
real networks. The conclusion revealed that link weight has
effects on network structures especially on the global properties.
Similar to the method used by Watt and Strogatz in their
discussion on small-world networks\cite{WS}, here we present a
more general method to investigate the effect of weight based on
regular networks and an idealized construction. The results are
interesting and valuable because real networks are usually
weighted and the redistribution of weight on links provides
another significant way to improve the structure and function of
networks.

\textbf{Effects of Weight on the Structure of Networks } Usually
there are two kinds of weight in weighted networks. One is
dissimilar weight, such as the distance between two airports.
Dissimilar weight has same meaning as the distance in binary
networks. The bigger is the weight, the larger is the distance
between two nodes. Another kind of weight is similar weight, such
as the times of coauthoring of scientists in scientific
collaboration networks and the flux in metabolic network. Similar
weight has opposite meaning as the general distance in binary
networks. The bigger weight means the smaller distance between two
vertices. If the weight is defined in sense of similarity, the
calculation of such similarity for example between two vertices
connected by two edges (with $w_{1}$ and $w_{2}$ respectively) is
$\tilde{w}=\frac{1}{\frac{1}{w_{1}}+\frac{1}{w_{2}}}$ \cite{Fan}.
In our following discussions, for simplicity and without losing
any generalization, the weight on links is dissimilar weight with
$w_{ij}\in[1,\infty)$ if not mentioned. Then the distance of a
path can be easily get from the sum of weight.

For weighted networks, the generalization of Watts-Strogatz
clustering coefficient is not entirely trivial. We must define a
quantity to measure the strength of connections within $i$'s
neighborhood. B. J. Kim has argued that this quantity should
fulfill four requirements and then provided a definition of
weighted clustering coefficients\cite{BJKim}. In order to apply
his definition to our weighted networks, we should first convert
our dissimilar weight into similar weight
$\tilde{w}_{ij}=\frac{1}{w_{ij}}$, then we have
$\tilde{w}_{ij}\in[0,1]$. Using the following equation
\begin{equation}
c_w(i)=\frac{\sum_{jk}\tilde{w}_{ij}\tilde{w}_{jk}\tilde{w}_{ki}}{\max_{ij}\tilde{w}_{ij}\sum_{jk}\tilde{w}_{ij}\tilde{w}_{ki}}
\end{equation}
we can get the local clustering coefficient for every node $i$ and
then get the average clustering coefficient of the network.

 In order to investigate the effects of weight on network structure,
we consider the same methodology of constructing WS small-world
networks. Instead of considering the link random rewiring
procedure, we study the effects of random redistribution of weight
on links for weighted regular network. Starting from a ring
lattice with $N$ vertices and $k$ edges per vertex, each edge has
a same weight {$w$=5} in the initial state. Firstly, we divide the
weight into a smaller unit $\Delta w$ ($\Delta w=1$). Secondly, we
extract randomly each unit with probability $p$. Lastly, we
equiprobably lay back each unit to all links. Without changing of
links among vertices, this construction allows us to have a
regular network with uniform weight distribution ($p$=0, $\delta$
distribution) and random weight distribution ($p$=1, Gaussian
distribution). And through the investigation of the intermediate
region $0<p<1$, we can know the effects of weight redistribution.

\begin{figure}
\center \includegraphics[width=8cm]{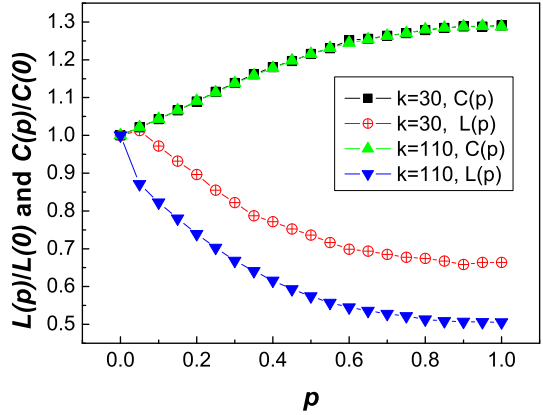}
\caption{Characteristic path length $L(p)/L(0)$ and clustering
coefficient $C(p)/C(0)$ for the family of randomly weight
redistributed networks ($N=200$). $k$ is the edges per vertex. All
the results (including Fig.2) are average over $20$ random
realizations of the redistribution process.} \label{SWNWeight}
\end{figure}

The same as WS small-world network, we use average path length
$L(p)$ and clustering coefficient $C(p)$ to quantify the
structural properties of the network. Fig. \ref{SWNWeight} reveals
that with the random redistribution of link weight, the average
path length decreases obviously, while the average clustering
coefficient increases. So it gives the similar phenomenon as
small-world effect, but here it is caused by the random
redistribution of weight instead of rewiring of links.

\begin{figure}
 \includegraphics[width=6.5cm]{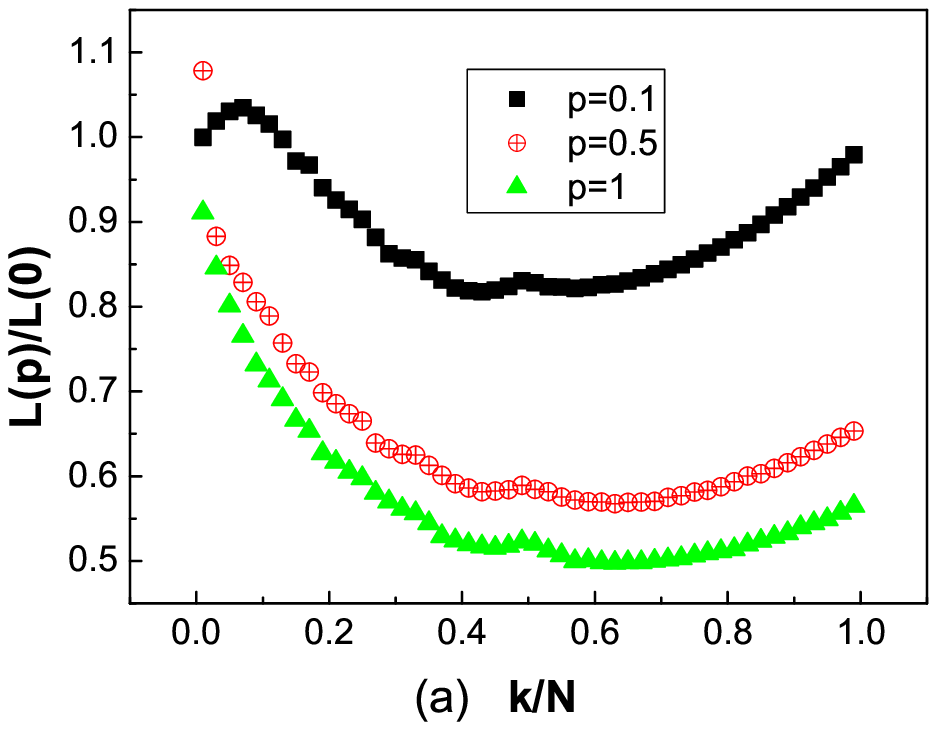}\includegraphics[width=6.5cm]{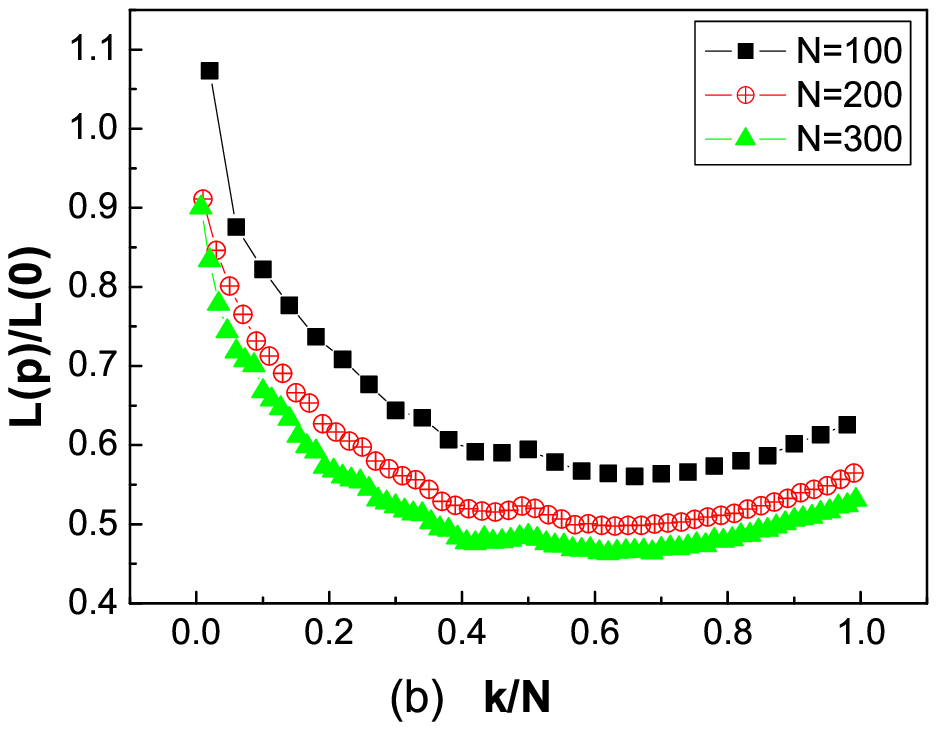}
\caption{(a) With a given random redistribution probability $p$,
$L(p)/L(0)$ as a function of network denseness described by
${k}/{N}$ ($N=200$). (b)For networks with different scale, the
curves are almost the same $(p=1)$. } \label{Npath}
\end{figure}

It seems that above effects of weight redistribution are more
significant in dense networks than in sparse networks. For a given
randomization probability $p$, Fig. \ref{Npath} (a) gives the
results of $L(p)$ as a function of degree $k$. With the increase
of the denseness of network, the decrease of $L(p)$ becomes more
and more obvious and $L(p)$ reaches an extremum to some extent.
Then the effect becomes less with the following increase of $k$.

It is interesting to notice that above effects are not related
with the scale of networks. We try this procedure for other
networks with different scales (number of vertices $N$). If we
re-scale the system with the quantity $k$/$N$ (which describes the
denseness of networks), we could get similar curves for
$L(p)/L(0)$. This demonstrates the robustness of the results (Fig
\ref{Npath}(b)).

\begin{figure}
 \includegraphics[width=6.5cm]{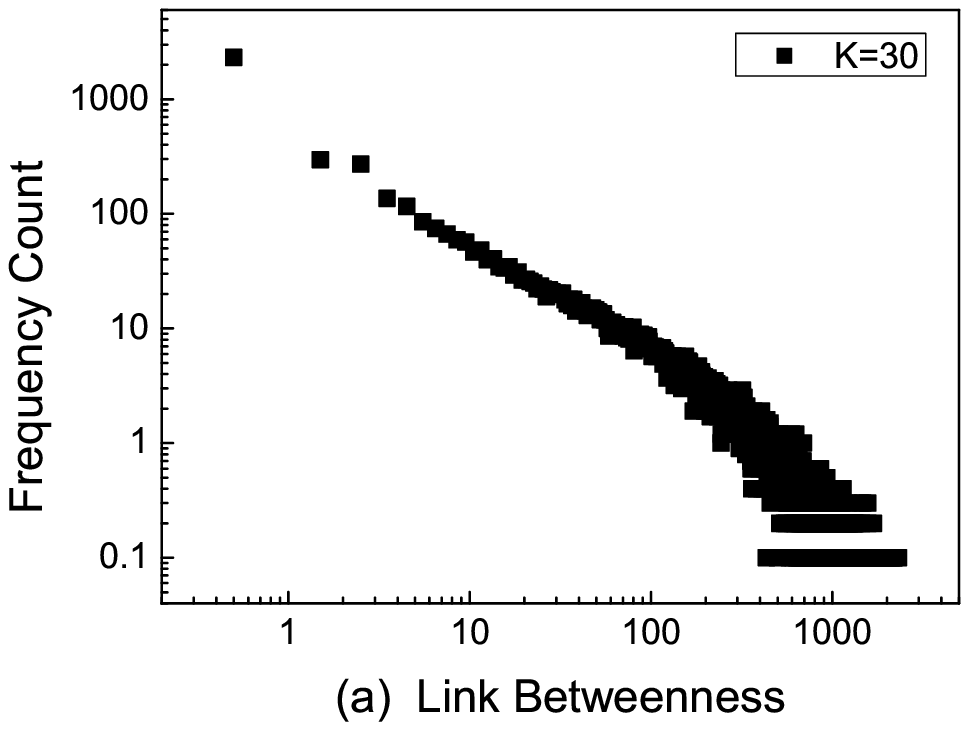}\includegraphics[width=6.5cm]{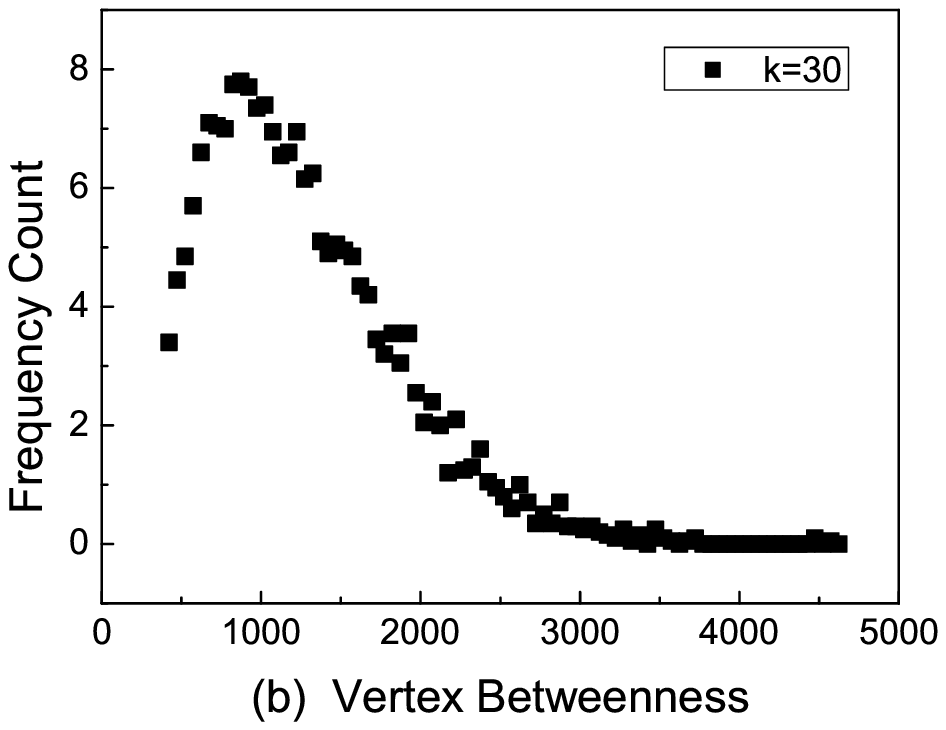}
\caption{ Distribution of link $(a)$ and node $(b)$ betweenness
after the full random redistribution of weight($p=1$).}
\label{betweenness}
\end{figure}

The idealized construction above gives the small-world phenomenon
due to the random redistribution of weight in networks. In order
to know the detailed change of network structure, we examine the
distribution of vertex and link betweenness after redistributing
weight. As shown in Fig. \ref{betweenness}, from the initial
homogeneous case, the distribution of link betweenness becomes
power-law and the distribution of vertex betweenness turns into
$\Gamma$ distribution. The result shows that some hubs emerge as a
result of the weight randomization. Hence the predicted changes in
the structure indicate that we have changed the global structural
features due to the only random redistribution of weight without
changing connections.

\textbf{Effects of weight randomization on Dynamics } For
investigating the functional significance of weight
redistribution, we impose this method on spin systems and
synchronization of chaotic systems on networks.

\textbf{A. Spin Systems} The phase transition of Ising model on
many kinds of networks have been studied deeply. In small-world
network, the transition temperature $T_c$ of Ising model increases
with the rewiring probability $p$ \cite{Tc1,Tc2}. In most previous
studies of spin model on networks, spin interactions have been
assumed to be uniform. But in reality, interactions are different
with each other. In reference \cite{DaunJeong}, the geometrical
distance is considered as the interaction strength between any two
nodes $i$ and $j$. In this letter, we consider both
nearest-neighbor and next nearest-neighbor
interactions\cite{Secondneibour} on weighted regular networks. The
Hamiltonian for Ising model on network is given by
\begin{equation}
H=-\frac{1}{2}\sum_{i\neq j}{J_{ij}\sigma _i\sigma _j},
\end{equation}
 where $\sigma _i(=\pm 1)$ is the Ising spin on node
$i$. Given the dissimilar weight $w_{ij}$ between any two nodes
$i$ and $j$ connected directly, the interaction $J_{ij}$ reads
$1/w_{ij}$ if $i$ and $j$ are connected directly, reads
$1/min_{s}(w_{is}+w_{sj})$  if $i$ and $j$ are next
nearest-neighbor and reads $0$ for other conditions. The
transition process is described  by the magnetization
$M=\frac{1}{N}\sum_{i=1}^{N}\sigma _i$. From a given temperature
$T$ and random initial spin state, we perform annealing algorithm
to describe the evolution of the system. In the process, if
$\Delta H= H_{new}-H_{old}>0$, transition probability from low
energy state to high energy state is $exp(\frac{-\Delta H}{T})$ .

\begin{figure}
 \center\includegraphics[width=8cm]{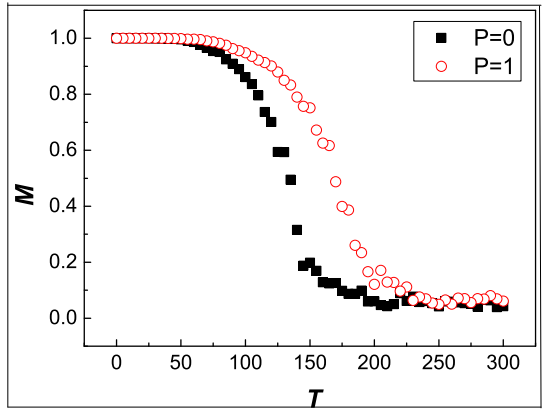} %\includegraphics[width=4.5cm]{M-P.eps}
\caption{Magnetization $M$ evolves with temperature $T$ for
homogeneous ($p=0$) and fully randomization ($p=1$) of weight
distribution. All the results are the average over 20 runs from
different random initial configurations of spins.} \label{Ising}
\end{figure}

Starting from a ring lattice with $N=500$ vertices and $k=200$
edges per vertex, each link has same weight {$w$=1} in the initial
network. Then we redistribute each $\Delta w=0.1$ randomly with
probability $p$. We find that the phase transition courses vary
with different redistribution probability $p$ (Fig. \ref{Ising}).
It shows that the randomization of weight has induced the increase
of critical temperature $T_c$. The redistribution of weight has
similar effects as the rewiring of links on phase transition.

\textbf{B. Synchronization of Chaotic System} Since their
introduction in $1989$ \cite{K.Kaneko}, coupled maps have been
studied as a paradigmatic example in the study of the emergent
behavior of complex systems as diverse as ecological networks, the
immune system, or neural and cellular networks. In recent years,
the synchronization of chaotic system on complex networks has
drawn much attention. In \cite{randomlycoupled}, chaotic system on
randomly coupled maps is investigated. It is found that the
synchronization properties of the system are strongly dependent on
the particular architecture. Graphs with the same number
connectivity might have very different collective behavior. The
previous works focused mainly on the effect of topology of network
on the synchronization. Here we investigate regular networks of
chaotic maps connected symmetrically and mainly focus the
influence of redistributing weight on synchronization. We take the
following coupled map
\begin{equation}
x_i(t+1)=(1-\varepsilon_{i})f(x_i(t))+\frac{\lambda}{m}\sum_{i\neq
j} J_{ij}f(x_j(t)),
\end{equation}
 where $x_i(t)$ is the state variable and $t$
denotes the discrete time. $f(x)$ prescribes the local dynamics,
and is chosen as the logistic map $f(x_i)=\alpha x_i(1-x_i)$ with
$\alpha=3.9$. $\varepsilon_{i}=\frac{\lambda}{m}\sum_{i\neq
j}J_{ij}$ gives the long-range coupling strength, where the sum is
taken over all the $m$ coupling nodes with $i$. We consider both
nearest-neighbor and next nearest-neighbor couplings and take the
same $J_{ij}$ as in Ising system and parameters are $N=300$,
$k=120$, $w=1$ and $\Delta w=0.1$. We make use of Degree of
Synchronization $d=\frac{1}{N}\sum_{i}^{N}|x_i-E(x_i)|$ to
discriminant whether to reach synchronization or not. In the
simulation, we neglect 500 steps as the transitional process and
$d$ is the average of following 1500 steps. All the results are
the average over 50 runs from different initial conditions.

\begin{figure}
 \includegraphics[width=6.5cm]{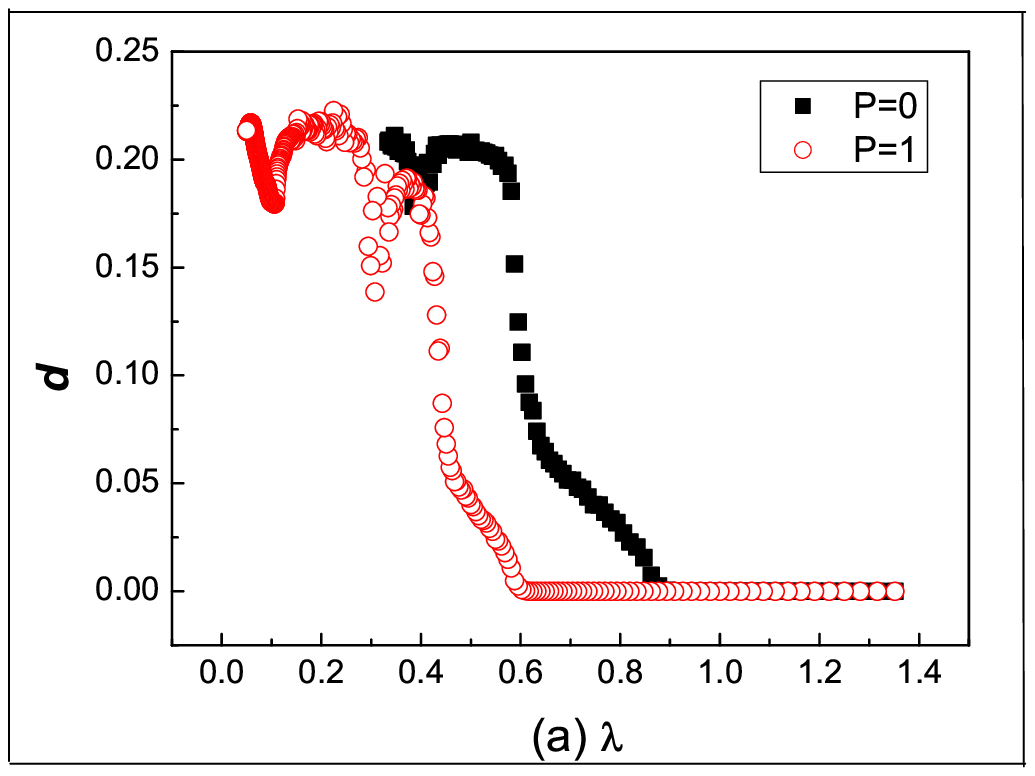}\includegraphics[width=6.5cm]{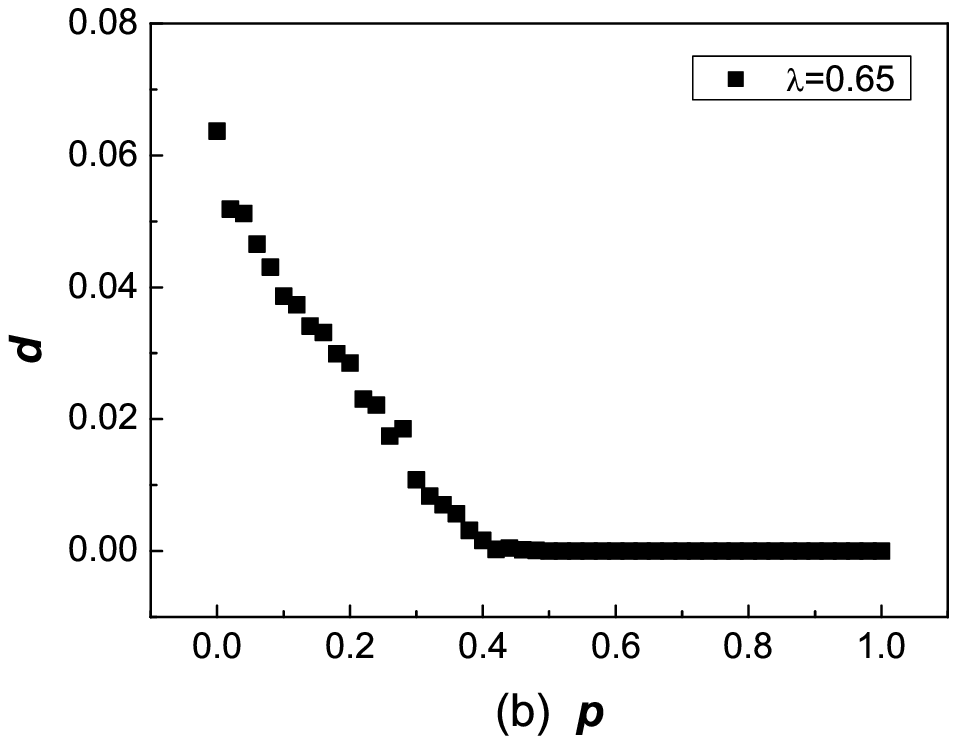}
\caption{(a) Final $d$ as the function of $\lambda$ for
homogeneous ($p=0$) and fully randomization ($p=1$) of weight
distribution. (b) The final $d$ as the function of probability $p$
with $\lambda =0.65$.}\label{synchronization}
\end{figure}

%Starting from a ring lattice with $N=300$ vertices and $k=120$
%edges per vertex, each link has a same weight($w=1$) in the
%initial network. Then we redistribute each $\Delta w$($=0.1$)
%randomly with probability $p$.
We first compare the final Degree of Synchronization $d$ of
homogeneous ($p=0$) and fully randomization ($p=1$) of weight
distribution for given average weight. The results
(Fig.\ref{synchronization} (a)) show that the redistribution of
weight is helpful to the synchronization of system.
Fig.\ref{synchronization} (b) gives the final $d$ as the function
of probability $p$. The randomization of weight enhances the
ability of synchronization.

In this letter, we emphasize the effects of weight on the
structural properties and function of networks. We explore a
simple model of networks with regular connections and homogeneous
weight. Instead of rewiring links, we introduce a method of
randomly redistributing weight of links. Its effects are
investigated both on structural properties and dynamical systems.
With the random redistribution of weight, we observed the similar
properties as in WS small-world networks. That is the average path
length declines while the clustering coefficient is increased. We
have also found that the random redistribution of weight can lead
to the increase of critical temperature in Ising model and enhance
the ability of synchronization of coupled chaotic systems.

Different from the most previous research on complex networks,
which focus on the topological structure and its influences to the
dynamical process, our investigation here illuminates the effects
of weight. From the previous study on dynamical systems on
networks with homogeneous coupling, it is already known that the
variation of link weight will affect the global function of
networks. But our results demonstrate that the change of weight
distribution can also cause some significant effects on the subtle
structural features and the functions of the given networks. These
results reveal that network topology coupled by weight
distribution should be essential to understand the structural
properties and function of weighted networks in real world.

The work is partially supported by 985 project and NSFC under the
grant No.70431002 and No.70471080.

\bibliography{apssamp}
Electronic address: yfan@bnu.edu.cn

\end{document}